\documentclass[11pt]{article}

\usepackage{amsmath,amsthm,amsfonts}

\usepackage{color,amsmath,amssymb,amsfonts}
\usepackage{epsf,epsfig,subfigure,psfrag,graphicx,afterpage}
\usepackage{url,lscape}
\usepackage[authoryear]{natbib}
\usepackage{multirow}
\usepackage{subfig}

\newcommand{\bftheta  }{\hbox{\boldmath$\theta$}}

\newcommand{\bfy}{\hbox{\boldmath$y$}}

\begin{document}

\title{A Conway-Maxwell-Poisson GARMA Model for Count Data}

\author{
  Ricardo S. Ehlers$^{\rm a}$\thanks{Corresponding author. Email: ehlers@icmc.usp.br}
  \vspace{6pt}\\
  $^{\rm b}${\em University of S\~ao Paulo, S\~ao Carlos, Brazil}
}

\date{}

\maketitle

\begin{abstract}

We propose a flexible model for count time series which has potential uses
for both underdispersed and overdispersed data. The model is based on
the Conway-Maxwell-Poisson (COM-Poisson) distribution with parameters
varying along time to take serial correlation into account. Model
estimation is challenging however and require the application of recently proposed methods to deal
with the intractable normalising constant as well as efficiently
sampling values from the COM-Poisson distribution.
\vskip ,3cm

\noindent Key words: Bayesian methods, COM-Poisson, Generalized ARMA, Intractable likelihood.

\end{abstract}

\section{Introduction}

Time series of counts are abundant in many areas of science and their
statistical analysis has received considerable attention. For example,
\cite{Zeger} and \cite{Chan} studied Poisson generalized linear models
where the mean follows a latent autoregressive process. 
Models for count time series embedded in the
framework of integer valued ARMA type models have also been discussed
in the literature (see for example \cite{Davis2} for an overview and also
\cite{biswas}).
Generalized autoregressive moving average (GARMA) models, introduced by 
\cite{Benjamin-etal03} and
extended in \cite{ehl2016c} for a Bayesian approach, have potential
uses for modeling overdispersed time series count data. In this paper,
we consider a more flexible model making use of the
Conway-Maxwell-Poisson (COM-Poisson) distribution which allows the
mean and the variance of count data to be modelled separetely and also
handles underdispersion (besides overdispersion).

Suppose the discrete random variable $Y$ represents count data with can present 
underdispersion, overdispersion, or equidispersion.
A flexible distribution 
to model such data is the Conway-Maxwell-Poisson (COM-Poisson, 
\cite{conway-maxell}) distribution 
for which the probability mass function is given by,
$$
p(y|\mu,\nu) = \left(\frac{\mu^{y}}{y!}\right)^{\nu}\frac{1}{Z(\mu,\nu)},
~\mu>0, \nu>0, ~y=0,1,2,\dots,
$$
where $Z(\mu,\nu)=\sum_{y=0}^{\infty}\left(\mu^{\nu}/y!\right)^{\nu}$
is an intractable normalising constant, having no closed form
representation for $\nu\ne 1$. This is the reparameterization proposed
by \cite{guikema}.
Clearly the Poisson distribution with parameter $\mu$ is a particular case when $\nu=1$,
in which case it exhibits equidispersion. Otherwise, the distribution will exhibit 
underdispersion ($\nu>1$) or overdispersion ($\nu<1$).

The mode of the distribution is given by $\lfloor{\mu}\rfloor$,
i.e. the largest integer below $\mu$ and there are two modes $\mu$ and
$\mu-1$ if $\mu$ is integer. Clearly, the moments of the distribution
are not available in closed form as a consequence of the intractable
normalising constant $Z(\mu,\nu)$. Approximations for the mean and the
variance are given by,
\begin{equation}\label{eq:E-Var}
E(Y)\approx\mu+\frac{1}{2\nu}-\frac{1}{2}
\quad\mbox{and}\quad 
Var(Y)\approx\frac{\mu}{\nu}
\end{equation}
which are based on an asymptotic representation of $Z(\mu,\nu)$
(\cite{minka-etal}, \cite{shmueli-etal}). 
If $\mu$ or $\nu$ (or both) are not small then these approximations
are quite accurate in which case $\mu$ closely approximates $E(Y)$.

Moments of the COM-Poisson can also be written in terms of derivatives of the
normalising constant and, as shown in \cite{minka-etal}, approximate
expressions are obtained from its asymptotic approximation,
\begin{eqnarray}\label{eq:moments}
E(Y) &=& \frac{\mu}{\nu}~\frac{\partial\log Z(\mu,v)}{\partial\mu}\nonumber\\
E[\log Y!] &=& -\frac{\partial\log Z(\mu,v)}{\partial\nu}
\approx \frac{1}{2\nu}\log\mu +\mu(\log\mu-1).
\end{eqnarray}

The main contributions in this paper are, to propose a COM-Poisson model with a 
Generalized ARMA (GARMA) structure for the mean of time series of counts as well as a
Bayesian approach to estimate parameters and compare models. This will
in turn require the application of recently proposed methods to deal
with the intractable normalising constant as well as efficiently
sampling values from the COM-Poisson distribution.
  
The remainder of this paper is organized as follows. In Section \ref{methods}, the
model adopted for a time series of counts and the methods used to estimate parameter
and make predictions are described. Examples of applications are provided in Section
\ref{applications} where an overdispersed time series of counts is analysed.
Section \ref{conclusion} concludes the paper.

\section{Model Description and Methods} \label{methods}

For a time series $y_1,\dots,y_n$ we assume that the conditional
probability mass function of each $y_t$ given the previous information
set 
$\mathcal{F}_{t-1}=\{y_1,\dots,y_{t-1},\mu_1,\dots,\mu_{t-1},\nu_1,\dots,\nu_{t-1}\}$
is given by, 
\begin{equation}\label{com}
  p(y_t|\mathcal{F}_{t-1}) = 
  \left(\frac{\mu_t^{y_t}}{y_t!}\right)^{\nu_t}\frac{1}{Z(\mu_t,\nu_t)}=
  \frac{q(y_t|\mathcal{F}_{t-1})}{Z(\mu_t,\nu_t)},
\end{equation}
where $q(y_t|\mathcal{F}_{t-1})$ is the unnormalised conditional 
probability of each $y_t$ assuming values $0,1,2,\dots$ and $Z(\mu_t,\nu_t)$ 
is the associated normalising constant. Since the COM-Poisson
distribution is a member of the exponential family we propose to
extend the idea in GARMA models introduced by \cite{Benjamin-etal03},
and assume the following linear predictor for the COM-Poisson model, 
\begin{equation*}
\log(\mu_t) = 
\sum_{j=1}^p\phi_j \log(y_{t-j}) + \sum_{j=1}^q\theta_j\{\log(y_{t-j}/\mu_{t-j})\},
\end{equation*}
thus accounting for the autocorrelation structure present in the data.
In practice, this expression is evaluated using $y_{t-j}^* =
\max(y_{t-j},c)$ with $0<c<1$ since each $y_t$ can assume a value zero.
In particular, if $\nu=1$ in (\ref{com}) this model reduces to the Poisson-GARMA
model as studied for example in \cite{ehl2016c}.
We also consider a relationship between the dispersion parameter
$\nu_t$ and lagged values of the time series using a logarithm link function,
\begin{equation*}
\log(\nu_t) = \sum_{j=1}^p\delta_j \log(y_{t-j}) 
\end{equation*}

Now let $\bftheta\in\mathbb{R}^d$ denote a $d$-dimensional
vector of parameters in the GARMA specification.
Then, the partial likelihood function for a COM-Poisson GARMA($p,q$) 
model is given by,
\begin{equation*}
  p(\bfy|\bftheta) = \prod_{t=r+1}^n p(y_t|\mathcal{F}_{t-1}) = \prod_{t=r+1}^n
  \left(\frac{\mu_t^{y_t}}{y_t!}\right)^{\nu_t}\frac{1}{Z(\mu_t,\nu_t)},
\end{equation*}
which clearly involves multiple intractable normalising constants. 
The log-likelihood is given by,
\begin{eqnarray*}
\log p(\bfy|\bftheta) &=& \sum_{t=r+1}^n \log p(y_t|\mathcal{F}_{t-1})\\
&=& \sum_{t=r+1}^n \nu_t y_t\log\mu_t -\sum_{t=r+1}^n\nu_t\log y_t! -\sum_{t=r+1}^n\log Z(\mu_t,\nu_t).
\end{eqnarray*}

In the Bayesian approach we also place a prior distribution on $\bftheta$
with density $p(\bftheta)$ and the, by Bayes theorem, the posterior distribution is given by,
$$
p(\bftheta|\bfy)\propto \prod_{t=r+1}^n p(y_t|\mathcal{F}_{t-1}) ~p(\bftheta),
$$
which is now a doubly-intractable posterior distribution since it depends
on integrating the right hand side which is itself intractable.
In terms of simulation, a direct application of the
Metropolis-Hastings algorithm would propose a candidate value
$\bftheta'\sim h(\cdot|\bftheta)$ and accept this move with probability, 
$$
\alpha(\bftheta,\bftheta')=\min\left\{1,
\frac{\prod_{t=r+1}^nq(y_t|\mathcal{F}_{t-1}')~p(\bftheta')~h(\bftheta |\bftheta')}
     {\prod_{t=r+1}^nq(y_t|\mathcal{F}_{t-1} )~p(\bftheta )~h(\bftheta'|\bftheta )}
~\frac{\prod_{t=r+1}^n Z(\mu_t,\nu_t)}{\prod_{t=r+1}^n Z(\mu_t',\nu_t')}\right\}.
$$
where,
\begin{equation*}
\log(\mu_t') = 
\sum_{j=1}^p\phi_j' \log(y_{t-j}) + \sum_{j=1}^q\theta_j'\{\log(y_{t-j}) - \log(\mu_{t-j}')\},
\end{equation*}
and
$\mathcal{F}_{t-1}'=\{y_1,\dots,y_{t-1},\mu_1',\dots,\mu_{t-1}'\}$.

It is therefore not computionally feasable to update the parameters $\bftheta$ using the 
standard Metropolis-Hastings algorithm. One way round this problem is to use the so 
called exchange algorithm proposed by \cite{murray-etal} which extends the algorithm in 
\cite{moller-etal} and basically augments the posterior state space with auxiliary data 
drawn from the likelihood. The idea is to include one further step by
proposing values 
$\bfy'=(y_1',\dots,y_n')$ given $\bftheta'$ and define the joint distribution,
$$
p(\bfy,\bfy',\bftheta,\bftheta')\propto
\prod_{t=r+1}^n \frac{q(y_t|\mathcal{F}_{t-1})}{Z(\mu_t,\nu_t)}
p(\bftheta) h(\bftheta'|\bftheta)
\prod_{t=r+1}^n \frac{q(y_t'|\mathcal{F}_{t-1}^*)}{Z(\mu_t',\nu_t')}.
$$
Note that the information set was redefine as
$\mathcal{F}_{t-1}^*=\{y_1',\dots,y_{t-1}',\mu_1',\dots,\mu_{t-1}'\}$
since each $y_t'$ is generate sequentially from the distribution of $y_t|\mathcal{F}_{t-1}^*$.
The acceptance probability now becomes,
$$
\alpha(\bftheta,\bftheta')=\min\left\{1,
\frac{\prod_{t=r+1}^nq(y_t |\mathcal{F}_{t-1}')~p(\bftheta')~h(\bftheta |\bftheta')}
     {\prod_{t=r+1}^nq(y_t |\mathcal{F}_{t-1} )~p(\bftheta )~h(\bftheta'|\bftheta )}
\frac{\prod_{t=r+1}^nq(y_t'|\mathcal{F}_{t-1} )}
     {\prod_{t=r+1}^nq(y_t'|\mathcal{F}_{t-1}^*)}
     \right\}.
$$
thus avoiding the need to evaluate a set of intractable normalising
constants which cancelled out. Note that the ratio of normalising constants
$Z(\mu_t,\nu)/Z(\mu_t',\nu_t')$ is actually being replaced by the
ratio of unnormalised probabilities $q(y_t'|\mathcal{F}_{t-1} )/q(y_t'|\mathcal{F}_{t-1}')$.

Since the exchange algorithm relies on sampling each auxiliary variable
$y_t'|\mathcal{F}_{t-1}^*$, $t=r+1,\dots,n$ from the COM-Poisson likelihood we
need this sampler to be computationally efficient. Also, it requires
exact draws from the likelihood given each parameter value and this is
the case for the sampler adopted in this paper and described in the next section.


\subsection{Sampling from the COM-Poisson Distribution}

Rewrite the probability mass function as,
$$
p(y|\mu,\nu)= q(y|\mu,\nu)/Z(\mu,\nu)
$$
where $Z(\mu,\nu)=\sum_{y=0}^{\infty}q(y|\mu,\nu)$ and consider an envelope density function
given by, 
$$
g(y)= q_g(y)/Z_g,
$$
where $Z_g=\sum_{y=0}^{\infty}q_g(y)$.
For a finite bounding constant $M$ the envelope inequality is given by
$Mg(y) > p(y|\mu,\nu)$. Then, a rejection sampling scheme would be to draw a value 
$y^*$ from $g(\cdot)$ and accept $y^*$ as a draw from $p(\cdot|\mu,\nu)$ 
with probability,
$$
\alpha(y^*) = \frac{p(y^*|\mu,\nu)}{M g(y^*)}.
$$

\noindent This is however clearly intractable if either $Z(\mu,\nu)$ or $Z_g$ 
is intractable. Likewise, the optimal value of the constant $M$ which is given by 
$\sup_y\{p(y|\mu,\nu)/g(y)\}$ is also impossible to obtain for intractable 
likelihoods.

Recently, \cite{Chanialidis2018} and \cite{benson-friel} proposed 
rejection sampling schemes to draw values from the COM-Poisson distribution 
without the need to evaluate the normalising constant. 
\cite{Chanialidis2018} proposed a rejection sampling scheme 
using an auxiliary
piecewise geometric distribution for which the normalising constant is
given in closed form.
\cite{benson-friel} proposed a considerably simpler approach which is
more useful in the context of time series models where we need to draw
each $y_t|\mathcal{F}_{t-1}$, $t=1,2\dots$. The scheme is briefly described below.

\subsubsection*{Upper Bound Approach}

In the rejection sampler, we note that the optimal value of the
bounding constant can be rewritten as,
\begin{eqnarray*}
M &=&\sup_y\left\{\frac{p(y|\mu,\nu)}{g(y)}\right\}
   = \frac{1/Z(\mu,\nu)}{1/Z_g}\sup_y\left\{\frac{q(y|\mu,\nu)}{q_g(y)}\right\}
   = \frac{Z_g}{Z(\mu,\nu)} B,
\end{eqnarray*}
where the constant $B=\sup_y\{q(y|\mu,\nu)/q_g(y)\}$ is tractable.
The acceptance probability is then given by,
$$
\alpha(y^*) = \frac{p(y^*|\mu,\nu)}{\frac{Z_g}{Z(\mu,\nu)} B ~g(y^*)}=
\frac{q(y^*|\mu,\nu)}{B ~q_g(y^*)}.
$$
\cite{benson-friel} then showed that we can sample from the COM-Poisson with parameters 
$\mu$ and $\nu$ by using a Poisson($\mu$) envelope with bounding constant,
$$
B= \left(\frac{\mu^{\lfloor\mu\rfloor}}{\lfloor\mu\rfloor !}\right)^{\nu-1}=
q(\lfloor\mu\rfloor,\mu,\nu-1),
$$
if $\nu\ge 1$ and a Geometric($p$) envelope with bounding constant,
$$
B= \left(\frac{\mu^{\lambda}}{\lambda!}\right)^{\nu} \frac{1}{(1-p)^{\lambda}p}=
q(\lambda,\mu,\nu)\frac{1}{(1-p)^{\lambda}p},
$$
if $\nu<1$ with $\lambda=\lfloor\mu/(1-p)^{1/\nu}\rfloor$. The value of $p$ 
in the geometric distribution should be chosen so as to maximise the acceptance 
rate of the sampler but this optimal value is infeasible to compute as it 
involves intractable normalising constants. \cite{benson-friel} proposed to 
chose $p$ by matching the geometric mean $(1-p)/p$ to the approximate COM-Poisson
mean $\mu+1/2\nu-1/2$ which results in $p=2\nu/(2\mu\nu+1+\nu)$. 
This is also the approach adopted here as it proved very efficient in
our applications. 

Finally, for the bounding constants above the acceptance probabilities for a 
proposed value $y^*$ are given by,
$$
\alpha(y^*)= \frac{q(y^*,\mu,\nu)}{q(\lfloor\mu\rfloor,\mu,\nu-1)q(y^*,\mu,1)}
$$
if $\nu\ge 1$ and,
$$
\alpha(y^*)= \frac{q(y^*,\mu,\nu)(1-p)^{\lambda}p}{q(\lambda,\mu,\nu)(1-p)^{y^*}p},
$$
if $\nu < 1$.

\subsection{Prior distributions}

The GARMA coefficients in the COM-Poisson GARMA model are defined in
the real line. Therefore, we assign independent normal prior
distributions centered about zero to these coefficients. The prior
variances $\sigma_{\phi}^{2}$, $\sigma_{\theta}^{2}$ and
$\sigma_{\delta}^{2}$ are chosen so as to reflect vague prior
knowledge on these coefficients.

\subsection{Bayesian Predictions}\label{sec:pred}

An important aspect of a time series model is its ability to predict future values of
the series. Given the observed time series $\bfy=\{y_1,\dots,y_n\}$,
the one-step ahead predictive probability mass function is given by,
\begin{eqnarray*}
  p(y_{n+1}|\bfy)= \int p(y_{n+1}|\bfy,\bftheta) p(\bftheta|\bfy)d\bftheta,
\end{eqnarray*}
which is not available in closed form. However, given a sample of size $N$, 
$\bftheta^{1},\dots,\bftheta^{N}$ from the posterior distribution of $\theta$
a Monte Carlo approximation is given by,
\begin{eqnarray}\label{cpred}
  \hat{p}(y_{n+1}|\bfy)= \frac{1}{N}\sum_{j=1}^N p(y_{n+1}|\bfy,\bftheta^{j}).
\end{eqnarray}
An approximation for the conditional probability mass function
$p(y_{n+1}|\bfy,\bftheta)$ in turn is obtained by generating
$y_{n+1}^{j,1},\dots,y_{n+1}^{j,L}$ from a COM-Poisson distribution 
with parameters $\mu_{n+1}^j$ and $\nu_{n+1}^j$ which are computed conditional
on $\bftheta^j$. This allows us to approximate the conditional
probabilities in (\ref{cpred}) by the sample proportions of
$y_{n+1}^j=k$, $k= 0,1,\dots$ and then calculate $\hat{p}(y_{n+1}|\bfy)$ accordingly. 

\section{Application} \label{applications}

\begin{figure}[h]\centering
\includegraphics{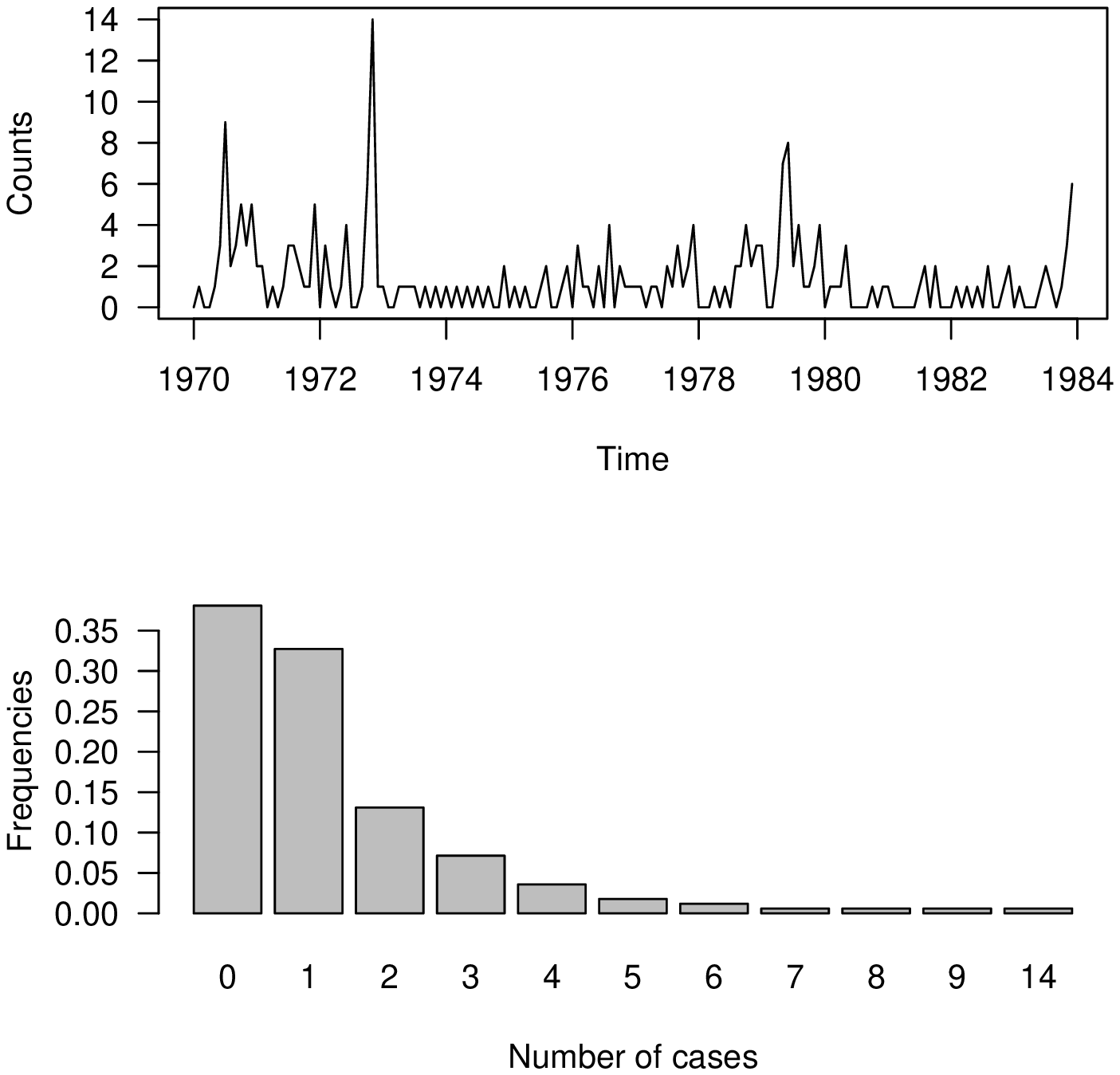}
\caption{The monthly number of cases of poliomyelitis in the United
  States between 1970 and 1983 and the associated frequencies.}
\label{fig:polio}
\end{figure}

The first real data set analysed is the time series with
the monthly number of cases of poliomyelitis in the United States between
1970 and 1983 (\cite{Zeger}). The data (198 observations) are depicted
as a time series in Figure \ref{fig:polio}. The mean and variance of
this series are $1.33$ and $3.5$ respectively which
together with the histogram in Figure \ref{fig:polio} clearly indicate
overdispersion. The following model was estimated,
\begin{eqnarray*}
  Y_t|\mathcal{F}_{t-1} &\sim& \mbox{COM-Poisson}(\mu_t,\nu_t),\\
  \log(\mu_t) &=& \phi \log(y_{t-1} + \theta\log(y_{t-1}/\mu_{t-1})\\
  \log(\nu_t) &=& \delta\log(y_{t-1}.
\end{eqnarray*}
for $t=1,\dots,n$.

The results shown in this section are based on running the proposed
sampler for 100,000 iterations, discarding the first 50,000 as burn-in
and skipping every 10th. This resulted in a final sample of 5000
values from the posterior distribution. The parameters were updated
jointly and a simple random walk
Metropolis was employed to propose new values using normal proposal
distributions centered about the current values with proposal
variances tuned to achieve an acceptance rate about 0.48.

\begin{figure}[h]\centering
  \includegraphics[height=0.6\textheight]{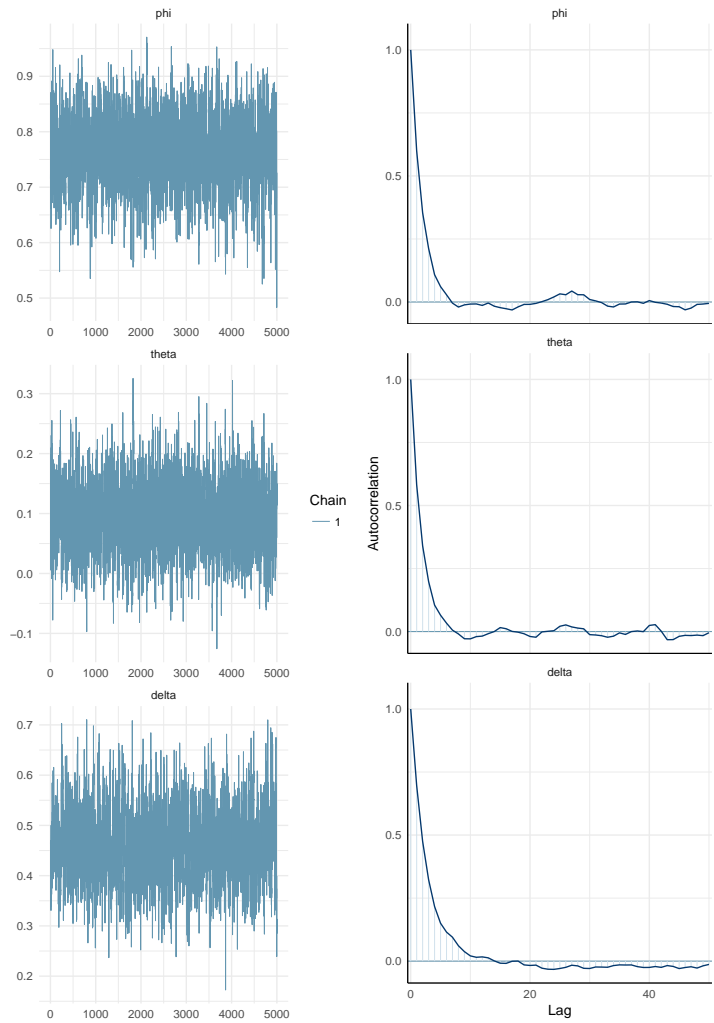}
  \caption{Trace and autocorrelation plots of the MCMC draws.}
  \label{fig:trace-acf}
\end{figure}

\begin{figure}[h]\centering
\includegraphics{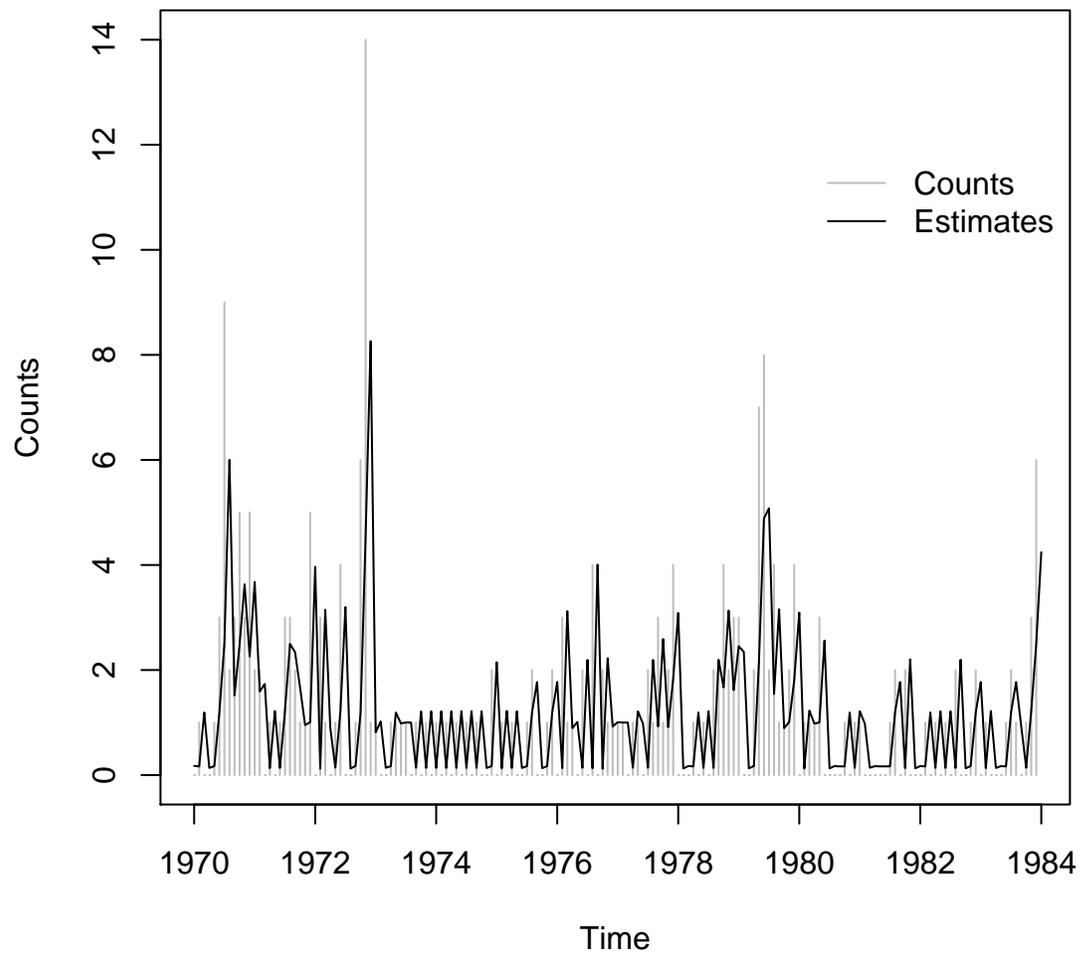}
\caption{Monthly number of cases of poliomyelitis in the United States between 1970
  and 1983 and estimated $\mu_t$.} 
\label{fig:data-mu}
\end{figure}

\begin{figure}[h]\centering
\includegraphics{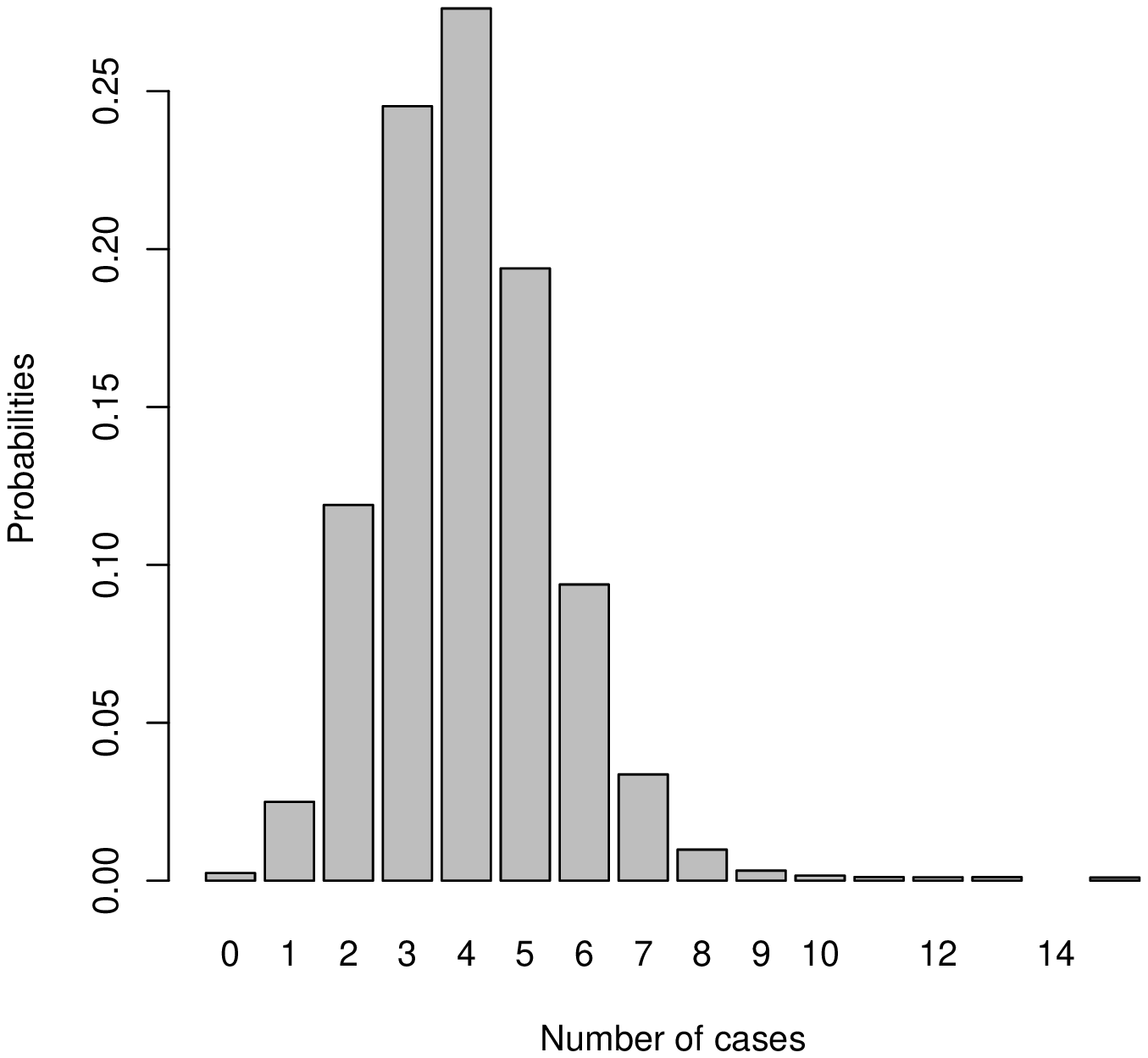}
\caption{Predictive one-step ahead distribution of the number of cases of poliomyelitis
  in the United States.}  \label{fig:hist1}
\end{figure}

Figure \ref{fig:trace-acf}
shows traces and the sample autocorrelations for the coefficients from
which we notice both good mixing in the parameter space and
autocorrelations vanishing fairly rapidly. Figure \ref{fig:data-mu}
shows the observed counts (as vertical lines) together with estimates
of $\mu_t$ given the sampled values of coefficients while Figure
\ref{fig:hist1} shows the estimated one-step ahead predicitive
distribution using the approach described in Section \ref{sec:pred}.

\section{Conclusions}\label{conclusion}

As pointed out in \cite{sellers-etal}, when dealing with count data
where the Poisson distribution plays an important role we should try and
extend the analysis to incorporate both overdispersion and
underdispersion features in the model. This is much so for time series of
counts where the degree of dispersion can additionally vary along time.

Based on advantages of using the COM-Poisson distribution to model
overdispersed and underdispersed data and recent advances on
computations for doubly intractable problems we proposed a COM-Poisson
GARMA model for time series of counts. Model parameters were estimated
using simulation based MCMC methods and the exchange algorithm coupled with a 
recently proposed clever way of generating values from a COM-Poisson
distribution. This COM-Poisson generator was also useful to obtain the
out-of-sample one-step ahead predictive distribution.

This is an ongoing work and the author is investingating ways to
perform model comparison which is challenging in the context of
intractable likelihoods.

\section*{Acknowledgments}

Ricardo Ehlers received support from S\~ao Paulo Research Foundation
(FAPESP) - Brazil, under grant number 2016/21137-2.


\end{document}